\documentclass[aps,dvips,showpacs]{revtex4}
\usepackage{graphicx}
\usepackage{graphics}
\usepackage{epsfig}
\usepackage{epsf}
\usepackage{subfigure}
\usepackage{latexsym}
\usepackage{amsmath}
\usepackage{amsfonts}
\usepackage[english]{babel}
\usepackage[latin1]{inputenc}
\def \p{\partial}
\def \half{\frac{1}{2}}
\def \sc{\mathop{\rm sc}\nolimits}
\def \sech{\mathop{\rm sech}\nolimits}
\def \dn{\mathop{\rm dn}\nolimits}
\def \sn{\mathop{\rm sn}\nolimits}

\def \arcsinh{\mathop{\rm arcsinh}\nolimits}

\def \bb{\bibitem}
\newcommand{\be}{\begin{equation}}
\newcommand{\ee}{\end{equation}}
\newcommand{\ba}[1]{\begin{array}{#1}}
\newcommand{\ea}{\end{array}}
\newcommand{\bea}[1]{\begin{equation}\left\{\begin{array}{#1}}
\newcommand{\eea}{\end{array}\right.\end{equation}}
\newcommand{\hs}[1]{\hspace{#1 mm}}

\begin{document}
\title{\bf Orbit-based deformation procedure for two-field models}
\author{V.I. Afonso$^a$, D. Bazeia$^a$, M.A. Gonzalez Leon$^b$, L. Losano$^a$, and J. Mateos Guilarte$^c$\\}

\affiliation{$^a$ Departamento de F\'\i sica, Universidade Federal da Para\'\i ba, Caixa Postal 5008,
                   58051-970, Jo\~ao Pessoa, Para\'\i ba, Brazil \\
$^{b}$ Departamento de Matematica Aplicada, Universidad de Salamanca, Spain\\
$^{c}$ Departamento de Fisica and IUFFyM, Universidad de Salamanca, Spain}

\vspace{3cm}

\begin{abstract}
We present a method for generating new deformed solutions
starting from systems of two real scalar fields for which
defect solutions and orbits are known. The procedure generalizes the approach introduced in a previous work [Phys. Rev. D {\bf66},
101701(R) (2002)], in which it is shown how to construct new models altogether with its defect solutions,
in terms of the original model and solutions.
As an illustration, we work out an explicit example in detail.
\end{abstract}
\maketitle
\section{Introduction}\label{intro}

Kinks, domain wall, vortices, strings and monopoles are all well known examples
of defect solutions with topological profiles for field theories in different dimensions \cite{Raj,Vil,Man}.
These solutions have been intensively studied since the seventies in high energy physics.
The interest has been continuously renewed in different branches of physics
since defect solutions in general appear in models of condensed matter as well as string theory.
In particular, systems of real scalar fields have attracted attention with very distinct
motivation, since they can be used to describe domain walls in supergravity \cite{sugra}
and braneworld scenarios with an extra dimension \cite{brane}.
Therefore, it is important to find explicit analytic solutions for this kind of systems and
is in this direction that the method conceived in ref. \cite{Ba1} has shown to be very useful.
In that work it was shown that knowing a defect solution of a scalar field model
for a single real field is enough for generating an infinity of new models with its solutions,
all written in terms of the original model and solutions.

For field theories involving two real scalar fields, the mathematical problem concerning
the integrability of the equations of motion is much harder, as one deals with a system of
two coupled second order nonlinear ordinary differential equations, and the configuration
space shows a distribution of minima that allows for a number of topological sectors.
One way of simplifying the problem is to consider potentials belonging to the (wide)
class corresponding to the bosonic sector of supersymmetric theories.
This kind of systems can be studied, via the introduction of a superpotential, in a first order formalism
which allows (stable) BPS configurations \cite{BPS}. Even in this case finding explicit solutions can be a highly non-trivial task and therefore, any method for obtaining new solutions would be of great utility.

For models with two interacting components, the solutions on each topological sector
determine orbits in the configuration space, which can be expressed as a constraint equation ${\cal O}(\chi_1,\chi_2)=0$.
Based on this fact it was introduced in \cite{Ra1} a procedure called trial orbit method
consisting in shooting an orbit and testing it on the equation corresponding to the model considered.
Later, this method was adapted \cite{Ba3} for the searching of BPS states of systems of
first order ODEs, leading to some advances.
In recent years, other general methods for the investigation of complicated nonlinear problems
arising in many-field systems which comprise multidefect solutions have been developed;
see, for instance, \cite{trodden,Ba4,dutra,gio,guilarte,vacha} and references therein.
Models of two scalar fields have also been used to describe complex phenomena such as the
entrapment of topological defects; see, for instance \cite{morris,ba,sut}.

As will be shown below, the field deformation method introduced in
\cite{Ba1} for one-field models also works for connecting ODE
systems of two first-order equations in models with two scalar
fields. The equations arising in this extended procedure are in
general much more complicated than their counterpart for a single
scalar field, given that not one but two deformation functions are
now required, and it happens to be difficult to realize which pair
of deformation functions would do the job in the right way,
generating a well behaved deformed potential and solutions
consistent with the equations of motion of the system.
To overcome this difficulty, we take into account the fact that the
actual solutions connecting two vacua of a topological sector live
restricted to orbits in field space. Therefore, by deforming
the first order equations for a two-field system while imposing the
orbit constraint, we assure the consistency of the solutions of the
deformed model at the level of the dynamical equations.

\bigskip
The paper is organized as follows. In the next Sec.~II we briefly review the deformation method for one component systems,
and we extend the deformation procedure to two-field models. In Secs.~III and IV we show how the incorporation of orbit constraints allows to obtain consistent deformed solutions for two-field models. Then we work out an explicit example in detail to illustrate the procedure in Sec.~V. We end this work with some comments and conclusions in Sec.~IV.

\section{The deformation method} \label{OneFieldDef}

We start with a system of a single real scalar field $\chi(x^\mu): {\mathbb R}^{1,1}\longrightarrow{\mathbb R}$, o a bi-dimensional Minkowski spacetime ($\mu=0,1$ with $x^0=x_0=t, \, x^1=-x_1=x$),
described by a Lagrangian with the usual form
\be
{\cal L} =\half \p_\mu\chi\p^\mu\chi - V(\chi),
\ee
\noindent where the potential $V(\chi)$ specifies the model.

For static configurations ($\chi=\chi(x):{\mathbb R}\longrightarrow{\mathbb R}$),
the equation of motion reads
\be
 \frac{d^2\chi}{dx^2} = \frac{d V(\chi)}{d\chi},
\ee
\noindent and the energy functional associated to the static solutions is given by
\begin{equation}\label{energy}
E[\chi]=\int dx \, \left[\frac{1}{2} \,\left(\frac{d\chi}{dx}\right)^2+V(\chi)\right]
\end{equation}

Requiring for the energy of the solutions to be finite results in the boundary conditions
$d\chi/dx \rightarrow 0$ and $V(\chi)\rightarrow 0$ as $x \rightarrow \pm \infty$.
Thus, the physical solutions are constant at infinity, and their asymptotic values are minima of the potential.
Performing a first integral of the equation of motion under these conditions,
we get to the first order equation
\be
 \left(\frac{d\chi}{dx}\right)^2 = 2V(\chi). \label{uno}
\ee

In this work we will restrict our study to solutions of this first order equation which present topological
(kink-like) character, in the sense that they connect two {\it different} minima of the potential.

It is convenient to consider potentials of the form $V(\chi)= \half [W'(\chi)]^2$,
where the prime means derivative with respect to the argument, and the functional $W(\chi)$ is the superpotential.
This allows us to write equation (\ref{uno}) as the gradient flow equations of $W$
\begin{equation}
\frac{d\chi}{dx}=\pm W'(\chi)\label{seis}
\end{equation}

Let us now describe the deformation procedure for a single real
scalar field model, as introduced in \cite{Ba1}. The prescription is
the following. First, we define the deformed potential as \be
U(\phi)=\frac{V(f(\phi))}{[f'(\phi)]^2} \label{dos}, \ee where $f$
is the {\it deformation function}.

This new potential determines the model for the deformed field $\phi$ through the deformed Lagrangian
\be
{\cal L}=\half\p_\mu\phi\,\p^\mu\phi-U(\phi).
\ee

The first-order ODE determining the defect profile in the deformed model reads
\begin{equation}  \label{FstOrdDef}
\left(\frac{d\phi}{dx}\right)^2=2\,U(\phi),
\end{equation}
with $U(\phi)$ given by (\ref{dos}). This can also be seen as the gradient flow equation of the {\it deformed}\hs{1}
superpotential ${\cal W(\phi)}$ defined by
\begin{equation}
\frac{d\phi}{dx}=\pm {\cal W}\,'(\phi)\quad ,\qquad
{\cal W}\,'(\phi)= \frac{W'(f(\phi))}{f'(\phi)}\label{FlowEqDef}
\end{equation}

\bigskip
Second, we connect solutions of the original and deformed models through the deformation function
\be\label{deformFunc}
\chi(x)=f[\phi(x)],
\ee

Here the deformation function $f(\phi)$ is assumed to be bijective - see however \cite{Ba4}.

Therefore, if defect solutions $\chi(x)$ of (\ref{uno}) are known, the link
between the two models provides the defect solutions $\phi(x)$ of
(\ref{FstOrdDef}) by just inverting the field transformation (\ref{deformFunc})
\begin{equation}
\phi(x)=f^{-1}(\chi(x)). \label{cuatro}
\end{equation}

\section{Extension to two-field models}

At the level of the first order equations, the deformation method can be
directly generalized to models with two scalar fields.
We start with the model
\begin{equation}
{\cal L}=\frac{1}{2} \partial_\mu
\vec{\chi}\cdot\partial^\mu\vec{\chi}-V(\vec{\chi}), \label{model}
\end{equation}
where $\vec{\chi}$ is an isospinorial real field
$\vec{\chi}=(\chi_1,\chi_2)$. Suppose that the potential energy
density can be written in the form \be V(\chi_1,\chi_2)=\half
\left(\frac{\p W}{\p \chi_1}\right)^2 + \half \left( \frac{\p
W}{\p\chi_2}\right)^2 \label{haja} \ee where the superpotential $W$
is a well behaved function in the space of scalar fields
$\vec{\chi}(x,t) \in{\rm Maps}({\mathbb R}^{1,1},{\mathbb R}^2)$. A
subtle point is the following: Because (\ref{haja}) is a PDE
equation there can be several independent solutions for $W$ - not
merely a global change of sign as in the one-field case - see
\cite{Gui2}. Then the static finite energy solutions (topological
defects) of this model satisfy the first-order ODE system
\be  \left\{ \ba{ccc} \label{ocho}
\frac{d\chi_1}{dx}&=&\frac{\partial W}{\partial \chi_1}\\\\
\frac{d\chi_2}{dx}&=&\frac{\partial W}{\partial \chi_2}
\ea\right.
\ee
\\
Now we choose a deformation function $\vec{f}:\mathbb{R}^2\to
\mathbb{R}^2$ such that
\begin{equation}
\vec{\chi}(x)=\vec{f}(\vec{\phi}(x))\quad \Leftrightarrow \quad
\left\{\begin{array}{ccc}
\chi_1(x)=f_1(\phi_1(x),\phi_2(x))\\\\
\chi_2(x)=f_2(\phi_1(x),\phi_2(x))\end{array}
\right., \label{nueve}
\end{equation}
where $\phi_1$ and $\phi_2$ are the deformed fields.
Then the first-order ODE system (\ref{ocho}) becomes
\begin{equation}
\left\{ \begin{array}{ccc}
\partial_1 f_1 \frac{d\phi_1}{dx}+ \partial_2 f_1 \frac{d\phi_2}{dx}&=& \partial_{f_1}W \\\\
\partial_1 f_2 \frac{d\phi_1}{dx}+ \partial_2 f_2 \frac{d\phi_2}{dx}&=& \partial_{f_2}W
\end{array}\right., \label{diez}
\end{equation}
\noindent where
\[
\partial_j f_i\equiv\frac{\partial f_i(\phi_1,\phi_2)}{\partial\phi_j}
\hspace{1cm}\mbox{and}\hspace{1cm}
\partial_{f_i}W\equiv\frac{\partial W(f_1(\phi_1,\phi_2),f_2(\phi_1,\phi_2))}{\partial f_i}.
\]

This system can be rewritten as
\begin{equation}
\left\{ \begin{array}{ccc}\frac{d\phi_1}{dx}&=&
\frac{1}{J(\vec{f})} \, \left( \partial_2f_2 \, \partial_{f_1} W-\partial_2 f_1 \, \partial_{f_2} W\right)\\ \\ \frac{d\phi_2}{dx}&=&
\frac{1}{J(\vec{f})} \, \left( \partial_1f_1 \, \partial_{f_2} W-\partial_1 f_2 \, \partial_{f_1} W\right)
\end{array}\right. \label{once}
\end{equation}
\medskip
\noindent where $J(\vec{f})=\partial_1 f_1 \,\partial_2f_2-\partial_2f_1\,\partial_1f_2$ (the Jacobian of $\vec{f}$) , and we assume that $J(\vec{f})\neq 0$. (Note that this fact
can be relaxed by restricting the deformation method to act on a
family of open sets in $\mathbb{R}^2$ where $J(\vec{f})\neq 0$).
Equations (\ref{once}) can be interpreted as the first-order ODE system
\begin{equation} \left\{ \begin{array}{ccc} \label{doce}
\frac{d\phi_1}{dx}&=&\partial_1{\cal W}\\\\
\frac{d\phi_2}{dx}&=&\partial_2{\cal W}
\end{array}\right.
\end{equation}
where ${\cal W}(\phi_1,\phi_2)$ is the superpotential of the deformed system and the right hand side derivatives denote $\partial_j{\cal W}=\frac{\partial {\cal W}(\phi_1,\phi_2)}{\partial \phi_j}$.

In terms of the original superpotential, the deformed one is determined by the PDE system
\begin{equation} \left\{ \begin{array}{ccc}
\frac{1}{J(\vec{f})} \, \left(\partial_2f_2 \, \partial_{f_1} W-\partial_2 f_1 \, \partial_{f_2} W\right)&=&
 \partial_1{\cal W}\\\\
\frac{1}{J(\vec{f})} \, \left( \partial_1f_1 \, \partial_{f_2} W-\partial_1 f_2 \, \partial_{f_1} W\right)&=&
\partial_2{\cal W}
\end{array}\right. \label{trece}
\end{equation}

Therefore, if any two-component defect solution
$\vec{\chi}(x)$
of (\ref{ocho}) is known, the link between the two models provides
two-component defect solutions
$\vec{\phi}(x)$
of (\ref{doce}) by calculating
\[
\vec{\phi}(x)=\vec{f}^{-1}(\vec{\chi}(x))
\]

Vice versa, if two-component defect solutions $\vec{\phi}(x)$ of
(\ref{doce}) are known, the link between both models provides two-component
defect solutions $\vec{\chi}(x)$ of (\ref{ocho}) by applying the transformation
\[
\vec{\chi}(x)=\vec{f}(\vec{\phi}(x))
\]

The existence of $\vec{f}^{-1}$ is associated to the above mentioned precisions
about the zeros of $J(\vec{f})$ via the Inverse Function Theorem.

\bigskip
At this point, the differences between working with one or more
scalar fields appear. First, the ODE system (\ref{doce}) is supposed
to give the solutions for the deformed system. This will be true
only when those solutions satisfy, besides the first order system,
the dynamical equations for the deformed potential \be {\cal
V}(\phi_1,\phi_2)=\frac{1}{2}[ \left(\partial_1{\cal W}\right)^2 +
\left(\partial_2{\cal W}\right)^2]. \ee

This is automatically satisfied by any solution of (\ref{doce}). But
beside this, one is assuming that the deformation leads to
well-behaved, smooth potentials. Therefore, we should ask for the
first order derivatives of the deformed superpotential ${\cal
W}(\phi_1,\phi_2)$ to be continue or, alternatively, for its
second order cross-derivatives to be identical $\,\p_{12}{\cal W}=\p_{21}{\cal
W}$. Imposing this condition on the systems (\ref{doce}) and
(\ref{trece}) leads to a very complicated constraint which suggest
no obvious choice of the deformation functions $f_1$ and $f_2$.
For this reason, in order to make progress we need to introduce some
assumptions to simplify the situation to a tractable case.
We will consider functions of the form $f_1=f_1(\phi_1)$ and $f_2=f_2(\phi_2)$.
This significantly reduces the complexity of the constraint,
which becomes the simple condition
\begin{equation} \label{cond1red}
\left((\p_{1}f_1)^2-(\p_{2}f_2)^2\right)\,\p_{f_1f_2}W \,\p_{1}f_1 \,\p_{2}f_2 =0
\end{equation}

This is true whenever $f_1$ and $f_2$ satisfy $\,\partial_1 f_1 = \pm\partial_2 f_2\,$
but, as $f_1(\phi_1)$ and  $f_2(\phi_2)$ are functions of different fields,
this expression seems to be nonsense.
However, since the solutions $\phi_1(x)$ and $\phi_2(x)$ live on an orbit
in configuration space, condition (\ref{cond1red}) must be understood as a
function of the solutions, therefore we rewrite it as
\begin{equation} \label{cond2}
\left[\frac{df_1(\phi_1)}{d\phi_1}= \pm\frac{df_2(\phi_2)}{d\phi_2}\right]_{\mbox{orbit}}
\end{equation}
When the deformation functions depend on one field only,
the PDE system (\ref{trece}) takes the simpler form
\bea{ccl}   \label{quinceb}
\p_1\,{\cal W}(\phi_1,\phi_2)&=&\displaystyle\frac{\p_{f_1} W[f_1,f_2]}{\partial_1 f_1}\\ \\
\p_2\,{\cal W}(\phi_1,\phi_2)&=&\displaystyle\frac{\p_{f_2} W[f_1,f_2]}{\partial_2 f_2}.
 \eea

\noindent Each one of these equations resembles the deformation recipe applied to the case
of a single scalar field model, so one could think that it is just a duplication of the one-field procedure.
However, the superpotential $W$ now depends on both deformation functions
$f_1(\phi_1)$ and $f_2(\phi_2)$, as a consequence of the interaction between the fields, and
then the deformed superpotential ${\cal W}$ depends on both
$\phi_1$ and $\phi_2$, and the resulting deformed model describes interacting fields.

As we have shown, requiring that the deformed model has a smooth potential
leads to a condition on the deformation functions.
Using this, and the fact that the fields are enforced to obey an orbit
we are able to construct a consistent deformation procedure,
detailed in the following section.

\section{ The orbit-based deformation}

Taking into account the considerations above, we present an orbit-based procedure for constructing
the deformation pair of functions.
The steps to be followed for deforming two-field interacting models are the following:
\begin{enumerate}

\item Choose a deformation function for one of the fields, for example a function $f_1(\phi_1)$.
Then, as stated in (\ref{nueve}) we define the deformed field by $\chi_1=f_1(\phi_1)$
(or, equivalently, $\phi_1=f_1^{-1}(\chi_1)$).
For the other field, $\chi_2=f_2(\phi_2)$ we can write
\begin{equation}\label{dieciseis}
 \phi_2 =\int \p_2 f_2^{-1}(\chi_2) d\chi_2= \int \frac{d\chi_2}{\p_2 f_2\left(f_2^{-1}(\chi_2) \right)}
\end{equation}

\item Choose the topological sector to be deformed and an orbit ${\cal O}(\chi_1,\chi_2)=0$, associated to this sector.
      Use this equation to write $\chi_1$ as a function of $\chi_2$, {\it i.e.}  $\chi_1 = F(\chi_2)$.

\item Then impose the condition (\ref{cond2}) on (\ref{dieciseis})
and use the expression $\chi_1=F(\chi_2)$ to obtain
\begin{equation} \label{diecisiete}
 \phi_2 =  \pm\int \frac{d\chi_2}{\p_1 f_1\left[f_1^{-1}(F(\chi_2))\right]}
        =  \pm\int \left[\p_1 f_1^{-1}(\chi_1)\right]_{\chi_1=F(\chi_2)} \,d\chi_2.
\end{equation}

\end{enumerate}

After integration we obtain $\phi_2$ as a function of $\chi_2$,
which is nothing but the inverse of the deformation function $f_2$.
This is the key result of the present work.

Thus, the chosen function $f_1(\phi_1)$ and the constructed one $f_2(\phi_2)$ form a pair
that takes the original model and solutions to a deformed model with a
smooth potential, and solutions satisfying the first order equations as well as the equations of motion.

Note that the procedure described above is orbit-dependent and the possibility of finding the second deformation function is restricted to the ability of explicitly integrating eq. (\ref{diecisiete}).

\section{A detailed example} \label{sec:BNRT}

As an example of application of the extended deformation method
we consider the model \cite{BNRT,Gui1}
\be\label{WBNRT}
W=\chi_1-\frac{1}{3}\chi_1^3 -r\chi_1\chi_2^2\;, \hs{15} r \in \mathbb{R}
\ee
\noindent It presents $4$ minima: $v_{AA}=[\pm1,0]$ on the $\chi_1$ axis, and $v_{BB}=[0,\pm\frac{1}{\sqrt{r}}]$ on the $\chi_2$ axis.

The corresponding first order system of equations is
\bea{ccl} \label{BNRTfstord}
\frac{d\chi_1}{dx}&=& 1-\chi_1^2-r\chi_2^2\\\\
\frac{d\chi_2}{dx}&=& -2r\chi_1\chi_2. \eea
The integrating factor for this system can be found
explicitly, and this allows to find the flow-line family of curves
\be\label{orbitBNRT}
r\chi_2^2-(1-2r)(1-\chi_1^2) + C
\chi_2^{\,\frac{1}{r}\,}=0, \qquad r\neq\frac12
\ee

\noindent where $C$ is an integration constant. Real values of  $C$
give orbits starting or ending in $v_{AA}$ minima, but there exist some
critical values. When $C=C^S=-2r(\pm\sqrt{r})^{\frac{1}{r}}$ the orbits start and end at
different axis, joining $v_{AA}$ and $v_{BB}$ minima. The existence
of these critical values determines several ranges of $C$ in
$\mathbb{R}$ for which the corresponding solutions of
(\ref{orbitBNRT}) are not kink orbits - see \cite{Gui1}.

\bigskip
So we have a two-field model and its general orbit equation depending on two parameters ($r$ and $C$). In order to apply the orbit-based procedure described in the previous section, we will consider separately the different kind of orbits, corresponding to different regions in parameter space.

\subsection{Elliptic orbit deformation}\label{Elliptic}
Let us first consider the simplest case in which the integration constant $C$ is taken to be zero and $r$ restricted to the interval $(\, 0,\half)$.
In this case, the orbits are ellipses and (\ref{orbitBNRT}) can be rewritten as
\be \label{veinticuatro}
\chi_2 = F(\chi_1)= \pm\sqrt{\frac{1-2r}{r}(1-\chi_1^2)} \hs{10} (\,C=0\,, \;\;\; 0<r< \mbox{$\half$}\,)
\ee

A two-field static solution for the system (\ref{BNRTfstord}), which satisfies this constraint
is \cite{BNRT}
\be
\chi_1(x)=\tanh(2rx), \hs{15}
\chi_2(x)=\sqrt{\frac{1-2r}{r}}\,\mbox{sech}(2rx).
\ee

\begin{figure}[ht]
\includegraphics[height=40mm]{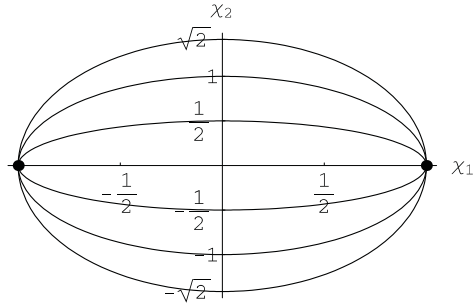}
\caption{Minima and orbits of the undeformed model for $C=0$ and
different $r$ values
($r=\frac{4}{9},\frac{1}{3},\frac{1}{4}$).}\label{fig1}
\end{figure}

In figure \ref{fig1} we show the vacua structure and some orbits of the model
for different values of the $r$ parameter.
While the two minima in the horizontal axis form a topological sector ($AA$-sector),
the two minima in the vertical axis ($BB$-sector) cannot be connected
by solutions of the first order system (\ref{BNRTfstord}).

Now, following the prescription established in the preceding section,
we choose a deformation function for $\chi_1$
\be \label{veinticinco}
\phi_1=f_1^{-1}(\chi_1)= {\rm arcsinh}(\chi_1)
\ee

For constructing the deformation function for the other field, $\chi_2$, we calculate the integral (\ref{diecisiete}) using (\ref{veinticuatro}) and (\ref{veinticinco})
\be
 \phi_2 = \int \frac{d\chi_2}{f_1'\left[f_1^{-1}(F(\chi_2))\right]}
        = \int \frac{d\chi_2}{\sqrt{2 - \left(\frac{r}{1-2r}\right)\chi_2^2}}
\ee
We obtain
\be
\phi_2= f_2^{-1}(\chi_2)= \sqrt{\frac{1-2r}{r}}\arcsin\left(\sqrt{\frac{r}{2(1-2r)}}\chi_2\right).
\ee
Thus
\be
f_2(\phi_2)=\sqrt{\frac{2(1-2r)}{r}}\sin\left(\sqrt{\frac{r}{1-2r}}\phi_2\right)
\ee
With the deformation functions at hand, and making use of (\ref{quinceb}),
we are able to write down the deformed potential, which reads
\be\ba{ccl}
U(\phi_1,\phi_2;r)&=& \frac{1}{2}\left[1-\sinh^2(\phi_1)-2(1-2r)\sin^2
\left(\sqrt{\frac{r}{1-2r}}\,\phi_2\right)\right]^2 {\rm sech}^2(\phi_1)\\\\
&& \hs{6}+
\,2\,r\,(1-2r)\sinh^2(\phi_1)\tan^2\left(\sqrt{\frac{r}{1-2r}}\,\phi_2\right)\label{ext}\ea\ee

Evaluating the fields $\chi_1$ and $\chi_2$ at the solution of the
original model we obtain a solitonic solution of the deformed
model specified by $U(\phi_1,\phi_2)$ \be \phi_1(x)={\rm
arcsinh}\left[\tanh(2rx)\right], \hs{15} \phi_2(x)=
\sqrt{\frac{1-2r}{r}}\arcsin\left(\frac{\sqrt{2}}{2}\,{\rm
sech}(2rx)\right) \ee

Condition (\ref{cond2}) is automatically satisfied by this solution.
This makes it consistent with the second order equations of the
deformed system, as can be explicitly verified.

\begin{figure}[hbtp]\centering
\mbox{\subfigure{\includegraphics[height=40mm]{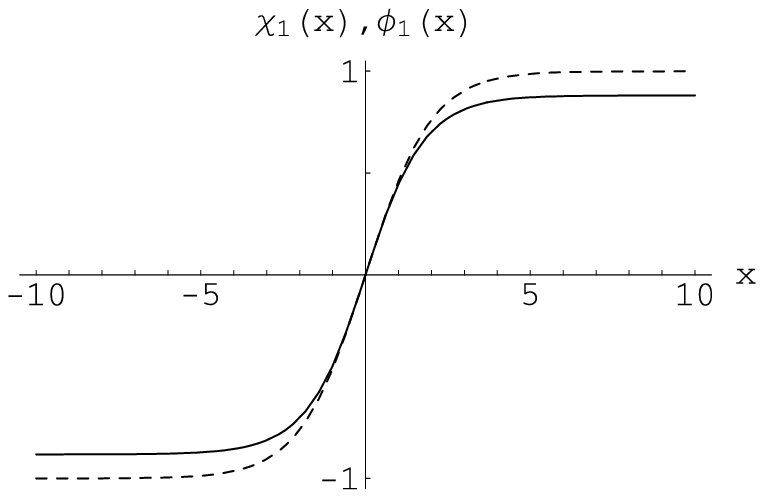}}\quad
\subfigure{\includegraphics[height=40mm]{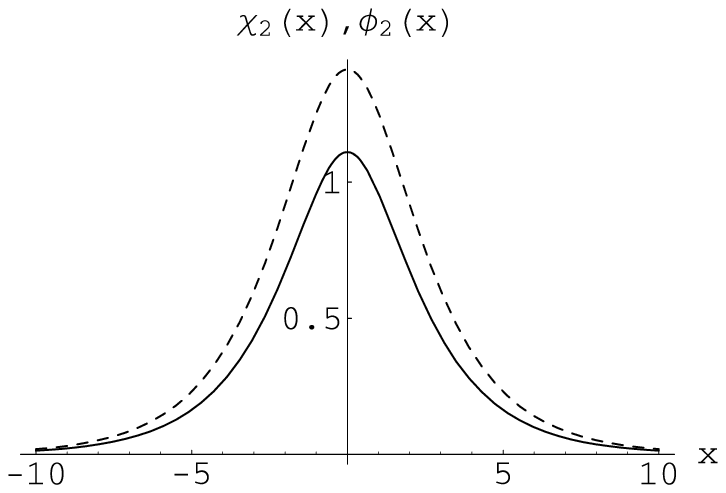}}
} \caption{Solutions of the undeformed (dotted line) and deformed
(solid line) models for $C=0$, $r=\frac14$.}\label{fig2}
\end{figure}

To show how this deformation acts, in figure \ref{fig2} we plot both, the deformed and the original solutions.

\begin{figure}[ht!]
\includegraphics[height=4cm]{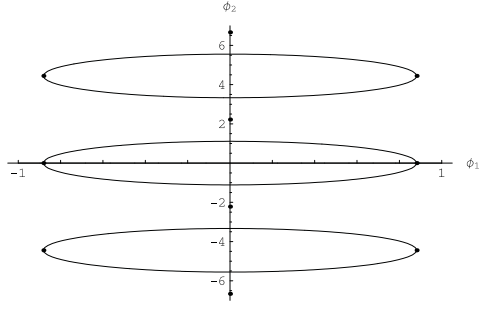}
\caption{Some vacua and orbits of the deformed model with $C=0$, $r = 1/4$.}\label{fig3}
\end{figure}

Substituting the original fields by its corresponding deformed partners in (\ref{orbitBNRT})
we obtain the deformed orbit
\be
\cosh^2(\phi_1)-2\cos^2\left(\sqrt{\frac{r}{1-2r}}\phi_2\right)=0.
\ee
which allows writing $\phi_2$ as a function of $\phi_1$
\be
\phi_2=\pm\sqrt{\frac{1-2r}{r}}\left[\arccos\left(\frac{\sqrt{2}}{2}\cosh(\phi_1)\right)+ k \pi \right] \hs{10} (\,k \in \mathbb Z\,)\; ,
\ee
and this explicitly shows that the new model presents a periodic vacua structure.
Such structure and some orbits of the deformed model are shown in fig. \ref{fig3}.

We can summarize the results obtained for the $AA$-sector as follows
\bea{lcl}
orb_{AA}&:&
r\chi_2^2-(1-2r)(1-\chi_1^2) =0; \hspace{2cm}  v_{AA}=\{[1, 0], [-1, 0]\}\\
\vec{\chi}(x)&=& \left(\,\tanh 2rx,\,\sqrt{\frac{1-2r}{r}}\,{\rm sech}\,2rx\,\right)\\
\vec{f}&=&\left(\,\sinh{\phi_1},\,\sqrt{\frac{2(1-2r)}{r}}\sin\left(\frac{\sqrt{r}}{\sqrt{1-2r}}\,\phi_2\right)\,\right)\medskip\\
orb_{AA\,def}&:&\cosh^2\phi_1-2\cos^2\left(\frac{\sqrt{r}}{\sqrt{1-2r}}\,\phi_2\right)=0.\\
v_{AA \,def}&=& \left[0,\sqrt{\frac{1-2r}{r}}\left({\rm arcsin}\frac{1}{\sqrt{2(1-r)}}+ k \pi \right)\right]\\
\vec{\phi}\,(x)&=& \left(\,{\rm arcsinh}\left(\tanh 2rx \right), \,\sqrt{\frac{1-2r}{r}}\arcsin\left(\frac{\sqrt{2}}{2}\,{\rm sech} 2rx\right)\,\right)\\
\eea

\subsection{Linear orbits deformation}

\begin{figure}[ht!]\centering
\includegraphics[height=40mm]{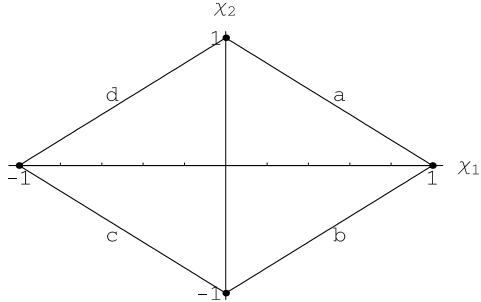}
\caption{$AB$-orbits ($C=\pm 2$, $r=1$).} \label{figABorbits}
\end{figure}

It is also possible to find explicit solutions for the BNRT model
(\ref{WBNRT}) in other regions in parameter space. For example, for
specific values of the integration constant $C$ and parameter $r$,
there are orbits connecting one minimun on the $\chi_1=0$ axis with
one on the $\chi_2=0$ axis ($AB$-sectors). As an illustration we
address now the $r=1$ case, and deform orbits for integration
constants $C=C^S=\pm 2$. There are four linear orbits, one for each
sector, that we label $a$, $b$, $c$ and $d$, starting from the first
quadrant and moving forward clockwise, as shown in figure
\ref{figABorbits}.

In order to deform these linear sectors we choose again the
deformation function for one of the fields as
$\chi_1=f_1(\phi_1)=\sinh(\phi_1)$, and construct the $f_2(\phi_2)$
by using the corresponding orbits. The resulting vacua structure and
orbits for the deformed models are shown below.

For the $a$ sector we obtain
\bea{lcl}
orb_{AB}^{(a)}&:&\chi_2+\chi_1-1=0 ; \hspace{2cm}  v_{AB}^{(a)}=\{[1, 0], [0, 1]\}\\
\vec{\chi}\,^{(a)}(x)&=& \left(\,\half(1-\tanh x),\,\half(1+\tanh x)\,\right)\\
\vec{f}\,^{(a)}&=&\left(\,\sinh{\phi_1},\,1+ \sinh{\phi_2} \,\right) \medskip\\
orb_{AB\,def}^{(a)}&:&\sinh{\phi_2}+\sinh{\phi_1}=0\\
v_{AB \,def}^{(a)}&=& \{[0, 0], [0,-\sinh^{-1}(2)],[\sinh^{-1}(1),\sinh^{-1}(-1)], [\sinh^{-1}(-1), \sinh^{-1}(-1)]\}\\
\vec{\phi}\,^{(a)}(x)&=& \left(\,-\sinh^{-1}\half(-1+\tanh x),\,\sinh^{-1}\half\left(-1+\tanh x\right)\,\right)\\
\eea
\begin{figure}[ht!]\centering
\mbox{\subfigure[Original
orbit]{\includegraphics[height=40mm]{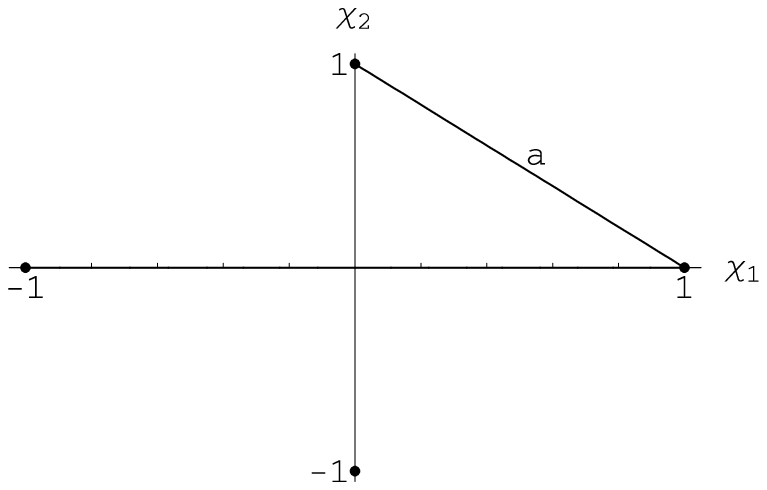}}
\quad \subfigure[Deformed
orbit]{\includegraphics[height=40mm]{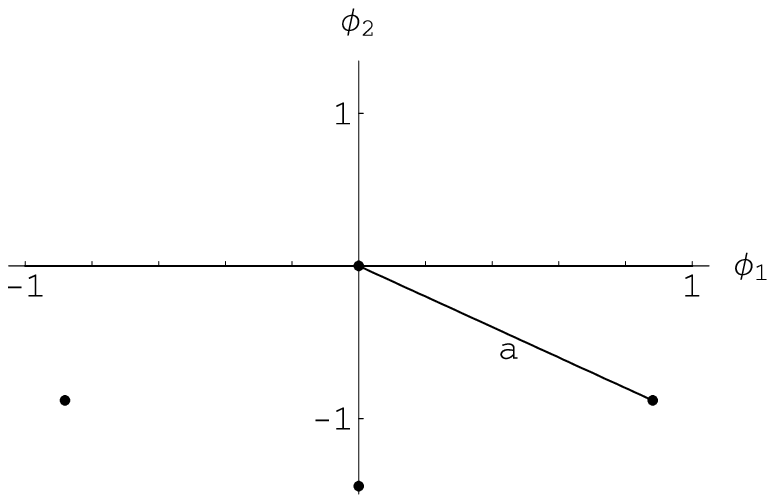}}}
\caption{Deformation of $AB^{(a)}$ sector ($C=-2$,
$r=1$).}\label{figABorbitsa}
\end{figure}
\begin{figure}[hbtp]\centering
\mbox{\subfigure{\includegraphics[height=35mm]{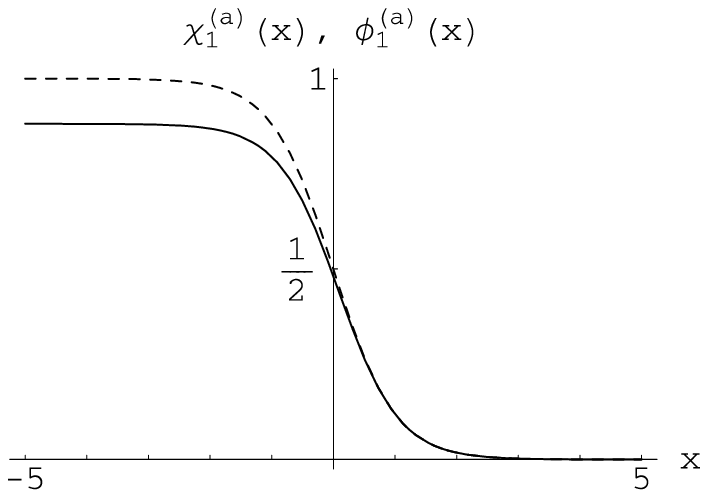}}
\quad
\subfigure{\includegraphics[height=35mm]{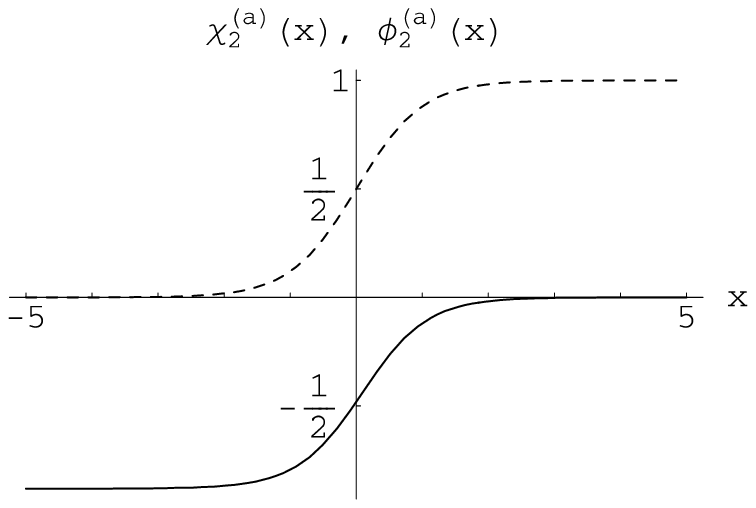}}
} \caption{Compared profiles of original (dashed line) and deformed
(solid line) defect solutions. $AB^{(a)}$ sector.}
\end{figure}

For the $b$ sector we get
\bea{lcl}
orb_{AB}^{(b)}&:&\chi_2-\chi_1+1=0 ; \hspace{2cm}  v_{AB}^{(b)}=\{[1, 0], [0,-1]\}\\
\vec{\chi}\,^{(b)}(x)&=& \left(\,\half(1-\tanh x),\,-\half(1+\tanh x)\,\right)\\
\vec{f}\,^{(b)}&=&\left(\,\sinh{\phi_1} ,\, -1+ \sinh{\phi_2}   \,\right) \medskip\\
orb_{AB\,def}^{(b)}&:&\sinh{\phi_2}-\sinh{\phi_1}=0\\
v_{AB \,def}^{(b)}& =& \{[0, 0], [0, \sinh^{-1} 2], [\sinh^{-1}(1), \sinh^{-1}(1)], [\sinh^{-1}(-1), \sinh^{-1}(1)]\} \\
\vec{\phi}\,^{(b)}(x)&=& \left(\,-\sinh^{-1}\half(-1+\tanh x),\,-\sinh^{-1}\half(-1+\tanh x)\,\right)\\
\eea
\begin{figure}[ht!]\centering
\mbox{\subfigure[Original
orbit]{\includegraphics[height=40mm]{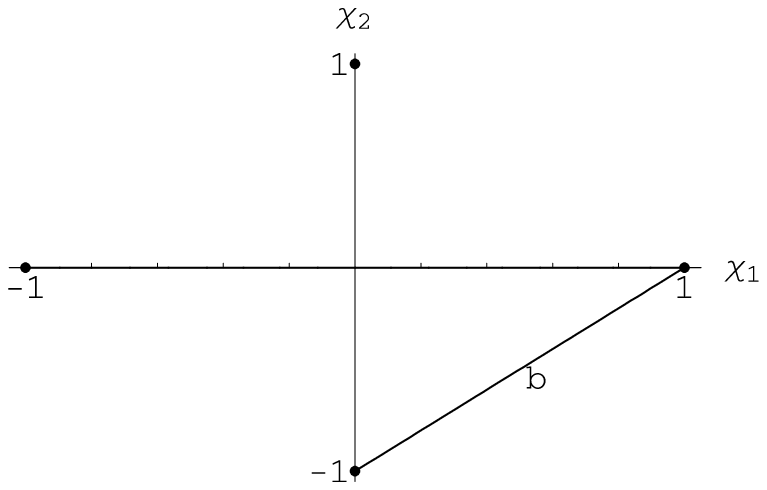}}
\quad \subfigure[Deformed
orbit]{\includegraphics[height=40mm]{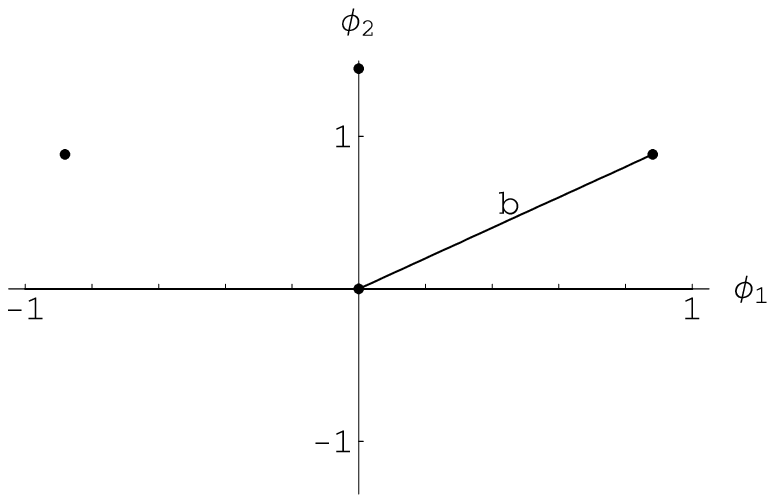}}}
\caption{Deformation of $AB^{(b)}$ sector ($C=2$,
$r=1$).}\label{figABorbitsb}
\end{figure}
\begin{figure}[hbtp]\centering
\mbox{\subfigure{\includegraphics[height=35mm]{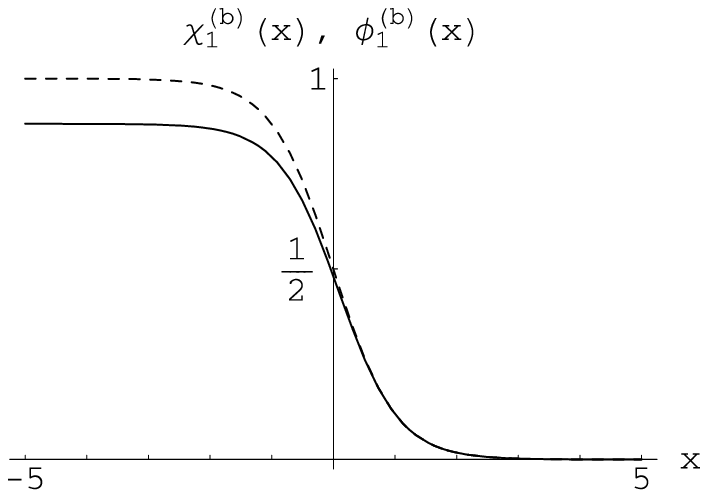}}
\quad
\subfigure{\includegraphics[height=35mm]{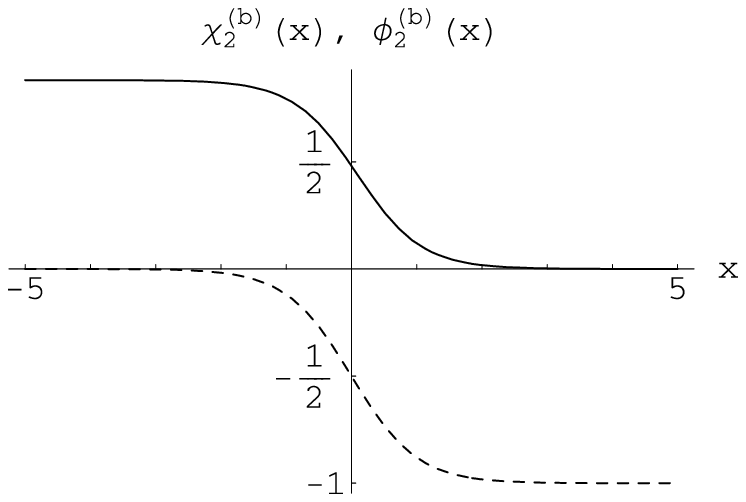}}
} \caption{Compared profiles of original (dashed) and deformed
(solid) defect solutions. $AB^{(b)}$ sector }
\end{figure}

In the case of the $c$ sector we obtain
\bea{lcl}
orb_{AB}^{(c)}&:&\chi_2+\chi_1+1=0 ;  \hspace{2cm} v_{AB}^{(c)}= \{[-1,0],[0,-1] \} \\
\vec{\chi}\,^{(c)}(x)&=& \left(\,-\half(1-\tanh x)   ,\,-\half(1+\tanh x)    ,\right)\\
\vec{f}\,^{(c)}&=&\left(\, \sinh{\phi_1}  ,\,  -1+ \sinh{\phi_2}  \,\right) \medskip\\
orb_{AB\,def}^{(c)}&:& \sinh{\phi_2}+\sinh{\phi_1}=0 \\
v_{AB \,def}^{(c)}& =& \{[0, 0], [0, \sinh^{-1} 2], [\sinh^{-1}(1), \sinh^{-1}(1)], [\sinh^{-1}(-1),\sinh^{-1}(1)]\} \\
\vec{\phi}\,^{(c)}(x)&=& \left(\, \sinh^{-1}\half\left(-1+\tanh x \right) ,\, -\sinh^{-1}\half\left(-1+\tanh x \right) \,\right)
\eea
\begin{figure}[ht!]\centering\mbox{
\subfigure[Original
orbit.]{\includegraphics[height=40mm]{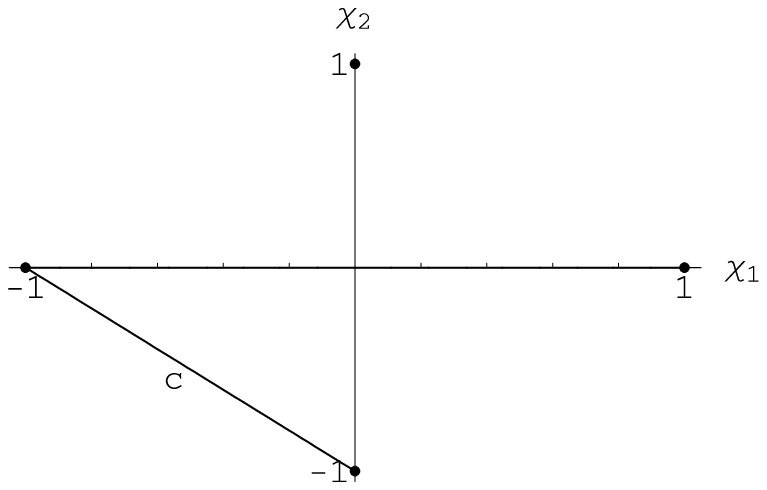}}
\quad \subfigure[Deformed
orbit.]{\includegraphics[height=40mm]{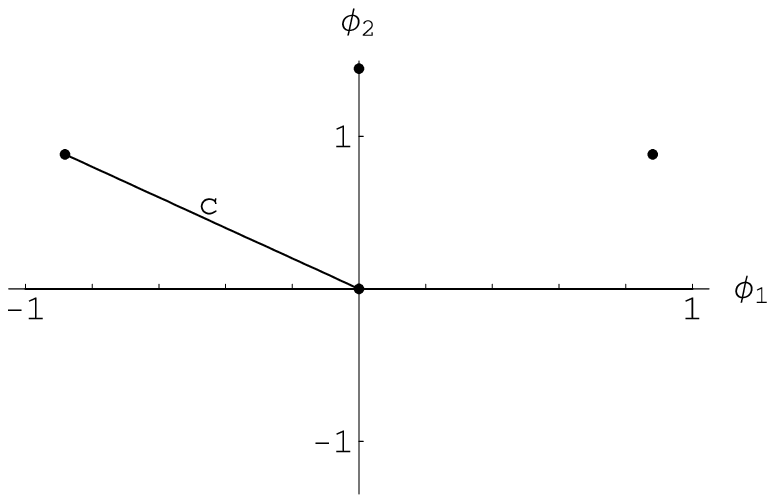}}}
\caption{Deformation of $AB^{(c)}$ sector ($C=2$,
$r=1$).}\label{figABorbitsc}
\end{figure}
\begin{figure}[hbtp]\centering
\mbox{\subfigure{\includegraphics[height=40mm]{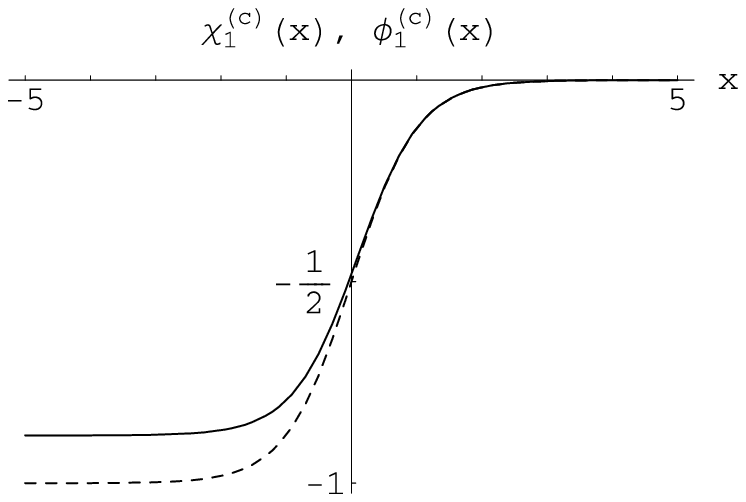}}
\quad
\subfigure{\includegraphics[height=40mm]{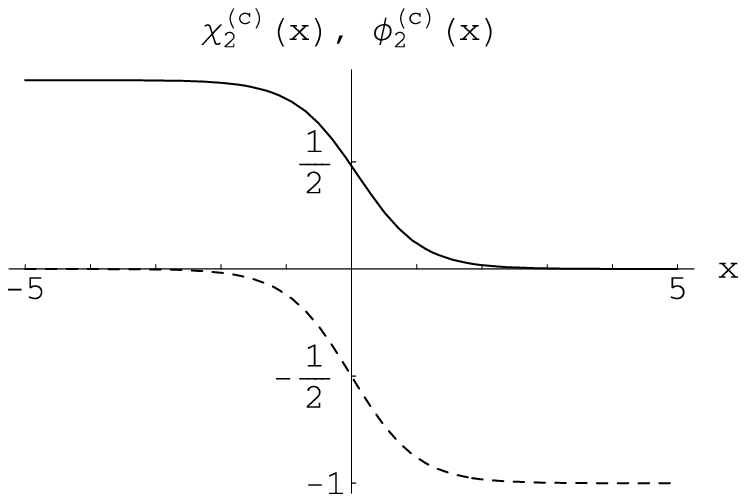}}
} \caption{Compared profiles of original (dashed line) and deformed
(solid line) defect solutions. $AB^{(c)}$
sector.}\label{figProfABorbc}
\end{figure}

Finally, for the $d$ sector we have
\bea{lcl}
orb_{AB}^{(d)}&:&\chi_2-\chi_1-1=0 ; \hspace{2cm}  v_{AB}^{(d)}=\{[-1,0],[0,1] \}\\
\vec{\chi}\,^{(d)}(x)&=& \left(\,-\half(1-\tanh x) ,\, \half(1+\tanh x),\right)\\
\vec{f}\,^{(d)}&=&\left(\,\sinh{\phi_1}  ,\, 1+ \sinh{\phi_2}\,\right) \medskip\\
orb_{AB\,def}^{(d)}&:&  \sinh{\phi_2}-\sinh{\phi_1}=0 \\
v_{AB \,def}^{(d)}& =& \{[0, 0], [0, -\sinh^{-1}2], [\sinh^{-1}(1),\sinh^{-1}(-1))], [\sinh^{-1}(-1),\sinh^{-1}(-1)]\} \\
\vec{\phi}\,^{(d)}(x)&=& \left(\,\sinh^{-1}\half\left(-1+\tanh x \right) ,\,
\sinh^{-1}\half\left(-1+ \tanh x \right) \,\right)\\
\eea
\begin{figure}[ht!]\centering
\mbox{\subfigure[Original
orbit.]{\includegraphics[height=40mm]{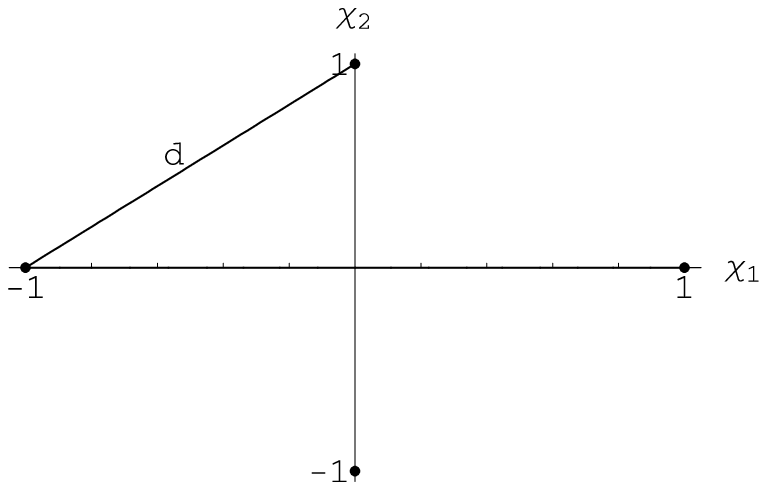}}
\quad \subfigure[Deformed
orbit.]{\includegraphics[height=40mm]{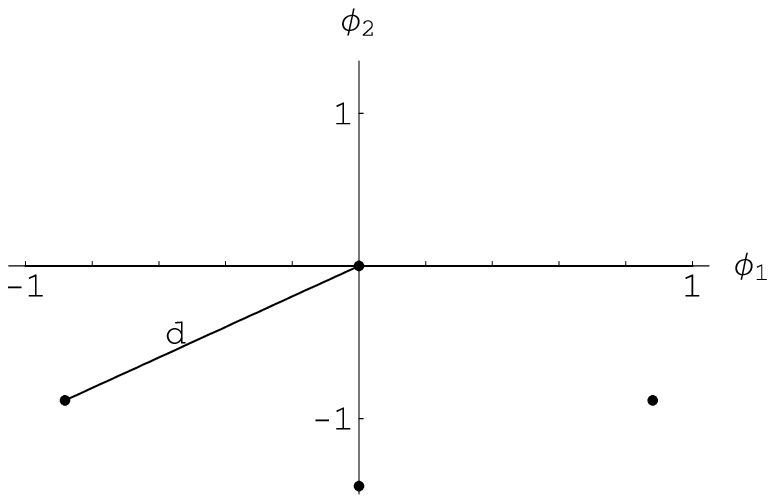}}}
\caption{Deformation of $AB^{(d)}$ sector ($C=-2$,
$r=1$).}\label{figABorbitsd}
\end{figure}
\begin{figure}[hbtp]\centering
\mbox{\subfigure{\includegraphics[height=40mm]{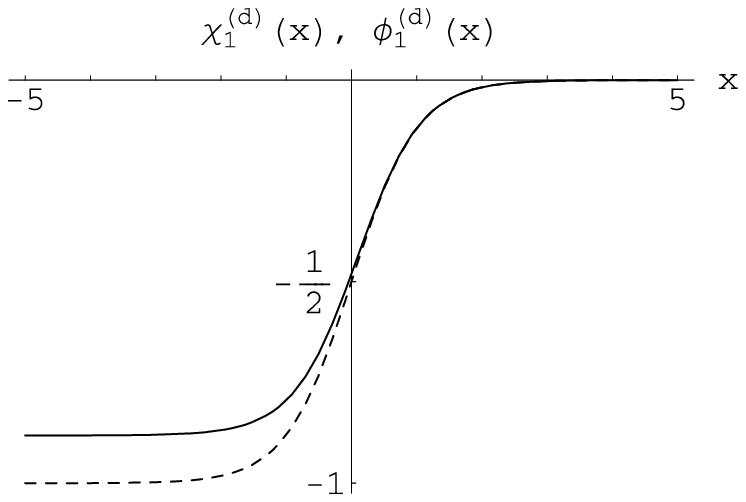}}
\quad
\subfigure{\includegraphics[height=40mm]{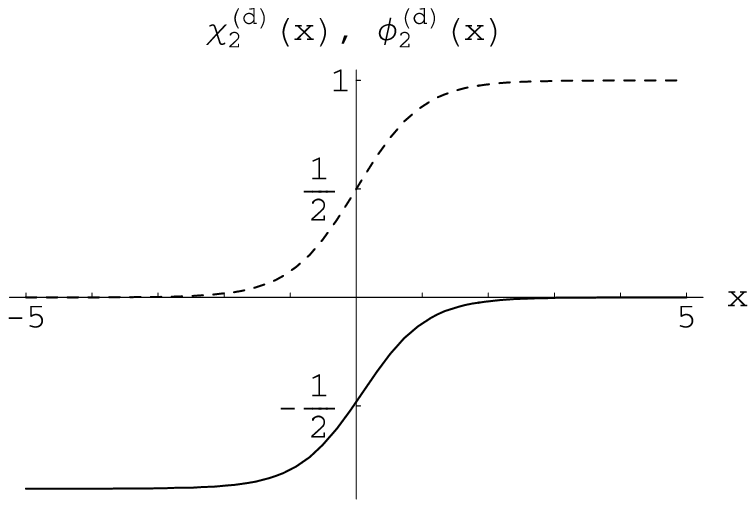}}
} \caption{Compared profiles of original (dashed line) and deformed
(solid line) defects. $AB^{(d)}$ sector.}
\end{figure}

In the above examples, we have considered two cases in which $C$ has been fixed.
Let us now move to other regions in parameter space, letting $C$ to be undetermined.
This will of course increment the complexity of the problem but, as we will show below,
there is still a lot of possibilities of generating new models.
This illustrates the richness of results that can be obtained by exploiting
the deformation method extended to interacting field models.

\subsection{An integrable case}
Recall the general orbit equation (\ref{orbitBNRT})
\begin{equation}\label{generalorbit}
\chi_1^2=1-\frac{r}{1-2r} \chi_2^2 -\frac{C}{1-2r}\,
\chi_2^{\frac{1}{r}}.
\end{equation}

\noindent Now, following the prescription for generating the deformed model, we are led to
\begin{equation}
\chi_1=f_1(\phi_1)=\sinh \phi_1; \quad\qquad
\phi_1=f_1^{-1}(\chi_1)=\arcsinh \chi_1\label{phi1C}
\end{equation}

and
\begin{equation}\label{phi2C}
\phi_2=\int \, \frac{d\chi_2}{\sqrt{2-\frac{r}{1-2r} \chi_2^2
-\frac{C}{1-2r}\, \chi_2^{\frac{1}{r}}}}
\end{equation}

\bigskip
Integral (\ref{phi2C}) is highly non trivial for arbitrary values of
$C$ and $r$. However it is integrable for some values of $r$, in
particular for $r=\frac{1}{4}$. For this value there is only one
critical constant $C^S=-\frac{1}{32}$, and kink orbits arise if $C
\in [C^S,\infty)$. In this case, (\ref{phi2C}) is an elliptic
integral, that can be written in the form \be\label{eq12}
\phi_2=\frac{1}{\sqrt{-2C}}\, \int
\frac{d\chi_2}{\sqrt{(\chi_2^2-\lambda_1)(\chi_2^2-\lambda_2)}}, \ee
with $\lambda_1=-\frac{1}{8C} \left( 1+\sqrt{1+64C}\right)$ and
$\lambda_2=-\frac{1}{8C} \left( 1-\sqrt{1+64C}\right).$

By an appropriate change of the integration variable, integral (\ref{eq12}) can be solved in terms of
Jacobian elliptic functions \cite{Abram}. The result is
\begin{equation}\label{eq15}
\phi_2=\frac{1}{\sqrt{\lambda_1}\sqrt{-2C}}\, \sn^{-1}\left(
\frac{\chi_2}{\sqrt{\lambda_2}};
\sqrt{\frac{\lambda_2}{\lambda_1}}\,\right).
\end{equation}
We note that for all the Jacobi elliptic functions appearing in this work we have to take its real part, as we are dealing with real scalar fields and they are solutions of a physical problem.

This solution presents distinguishable behaviors depending on $C$ taking values on the regions $(-\frac{1}{32},-\frac{1}{64})$, $(-\frac{1}{64},0)$ or $( 0,\infty)$.
For the special values $C=0$ and $C=-\frac{1}{64}$, the deformed system presents field solutions in terms of elementary functions rather than Jacobi elliptic functions.

Putting the original field solutions in terms of the deformed ones
\begin{equation}\label{eq15b}
\chi_1=\sinh \phi_1\; ,\qquad
\chi_2=\sqrt{\lambda_2} \, \sn \left(\sqrt{\lambda_1}\sqrt{-2 C}\, \phi_2; \sqrt{\frac{\lambda_2}{\lambda_1}}\right),
\end{equation}
we can write the explicit form the deformed model potential. We obtain
\begin{eqnarray}
U(\phi_1,\phi_2; \frac{1}{4} ,C)&=&\frac{1}{2} \left[ 1-\sinh^2
\phi_1 -\lambda_2 \sn^2 \left( \sqrt{\lambda_1}\sqrt{-2C} \, \phi_2;
\sqrt{\lambda_2/\lambda_1}\right)\right]^2 \sech^2\phi_1\nonumber\\
&& -\frac{\lambda_1 C}{2} \sinh^2\phi_1 \, \frac{ \sc^2
\left(\sqrt{\lambda_1}\sqrt{-2C} \, \phi_2;
\sqrt{\lambda_2/\lambda_1}\right)}{\dn^2 \left(
\sqrt{\lambda_1}\sqrt{-2C} \, \phi_2;
\sqrt{\lambda_2/\lambda_1}\right)}\label{15c}
\end{eqnarray}

A field solution of the undeformed system (\ref{BNRTfstord}),
satisfying the orbit (\ref{generalorbit}) for $r=\frac{1}{4}$ is \be
\chi_1(x)=\frac{\sinh(x)}{\cosh(x)+b^2},\hs{15}
\chi_2(x)=2\frac{b}{\sqrt{\cosh(x)+b^2}}, \ee where
$b^2=1/\sqrt{1+32\,C}$.
Therefore, using the deformation functions (\ref{eq15b}), the solutions for the deformed system read
\be
\phi_1(x)=\arcsinh\left[\frac{\sinh(x)}{b^2+\cosh(x)}\right], \hs{7}
\phi_2(x)=\frac{1}{\sqrt{-2\,\lambda_1C}}\,\sn^{-1}\left[2\frac{b/\sqrt{\lambda_2}}{\sqrt{\cosh(x)+b^2}};
\sqrt{\frac{\lambda_2}{\lambda_1}}\right]\label{phi1phi2}
\ee

\begin{figure}[hbtp]\centering
\mbox{\subfigure{\includegraphics[height=40mm]{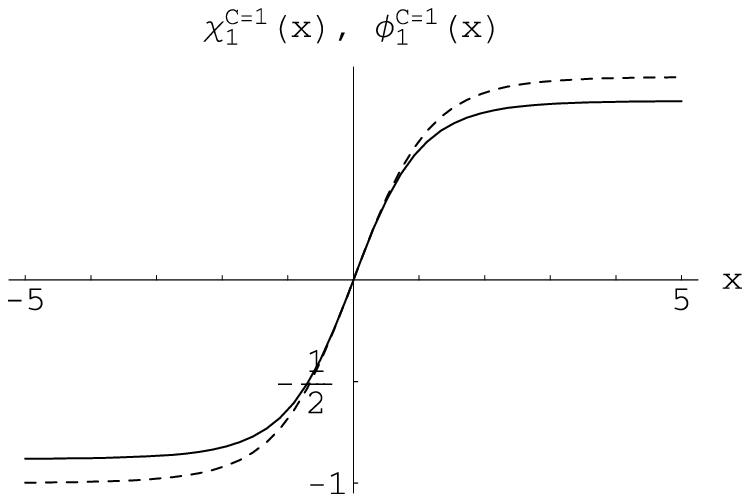}}
\quad
\subfigure{\includegraphics[height=40mm]{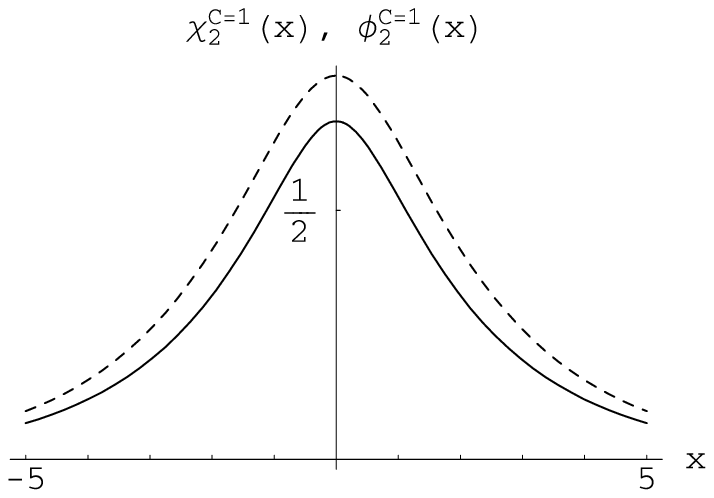}}
} \caption{Deformed (solid) and undeformed (dotted) solutions for
BNRT model ($r=\frac{1}{4}$, $C=1$).}\label{figMath}
\end{figure}

In figure \ref{figMath} we plot the original and deformed profiles
of the field solution (taking $C=1$). It is remarkable how similar
is the behavior of these solutions and the ones plotted in figure
\ref{fig2}, despite the much more involved analytical expressions in
this last case (compare formula (\ref{15c}) with formula
(\ref{ext})).

\bigskip

From the results obtained above we stress that the orbit-based deformations
can be applied to every kink orbit of the BNRT model (\ref{WBNRT})
for the integrable case of $r=1/4$.
For almost any value of the constant $C$ we obtain different models
in terms of  elliptic Jacobi functions, with two exceptions: $C=0$ and $C=-1/64$.
Also, we note that there is only one kink solution for each member of the family of deformed models,
coming from the appropriate kink orbit in the original model.

\subsection{Another integrable case.}

For $r=1$ the deformed model for every kink orbit can also be found,
given that, as for the former case, it is possible to integrate (\ref{phi2C}) for any value of $C$.
In fact, setting $r=1$ in (\ref{phi2C}) and taking $\phi_1=\arcsinh(\chi_1)$, we obtain the deformed $\phi_2$ field
\be
\phi_2=\ln\left(\half C+\chi_2+\sqrt{2+C\chi_2+\chi_2^2}\right)
\ee

Writing the original field solutions in terms of the deformed ones,
we can find the explicit form for the deformed model potential, which turns out to be
\be\ba{rcl}\label{defPot-r1}
U(\phi_1,\phi_2; r=1, C)&=&\half\sech(\phi_1)^2\left[1-\sinh(\phi_1)^2-
\frac{1}{64}\,e^{-2\phi_2}(-8 + 4\,e^{2\phi_2} - 4\,C\,e^{\phi_2}+C^2)^2\right]^2\\
&&+\frac{1}{4}\sinh(\phi_1)^2
e^{-2\phi_2}\left[\frac{-8+4\,e^{2\phi_2}- 4\,C\,e^{\phi_2}+C^2}{8\,
e^{-\phi_2}+4\,e^{\phi_2}-C^2\,e^{-\phi_2}}\right]^2 \ea\ee

The solutions of the BNRT model for $r=1$ can be written as
\be\label{solsr1} \chi_1=
\frac{(e^{2x})^2-C_1^2+C_2^2}{(e^{2x}-C_1)^2-C_2^2}, \hs{10} \chi_2
=-\frac{2 C_2 e^{2x}}{(e^{2x}-C_1)^2- C_2^2} \ee with $C_2=2\,
C_1/C$ and $\left|C\right|\geq \left|C^S\right|=2$. As was said
before, the solutions given in formula (\ref{solsr1}) are not kinks
if $|C|<2$, as both $\chi_1$ and $\chi_2$ become singular for this values of $C$.

The corresponding deformed solutions read
\be\ba{rcl}\label{defsolsr1}
\phi_1&=&\arcsinh\left(\frac{(e^{2x})^2-C_1^2+C_2^2}{(e^{2x}-C_1)^2-C_2^2}\right)\\\\
\phi_2&=& \ln\left[\half C-\frac{2 C_2 e^{2x}}{(e^{2x}-C_1)^2-C_2^2}
+\sqrt{2+\frac{4 C_2^2 (e^{2x})^2}{(e^{2x}-C_1)^2-C_2^2)^2} - \frac{4 C_1 e^{2x}}{(e^{2x}-C_1)^2-C_2^2}}\,\,\right]
\ea\ee

\medskip
For the specific value $C=2$, the solutions (\ref{solsr1}) are both kink-like, as well as the deformed solutions (\ref{defsolsr1}), and its profiles are very similar to that of the case $C=-2$, plotted in Fig. \ref{figProfABorbc}.

\begin{figure}[hbtp]\centering
\mbox{\subfigure{\includegraphics[height=40mm]{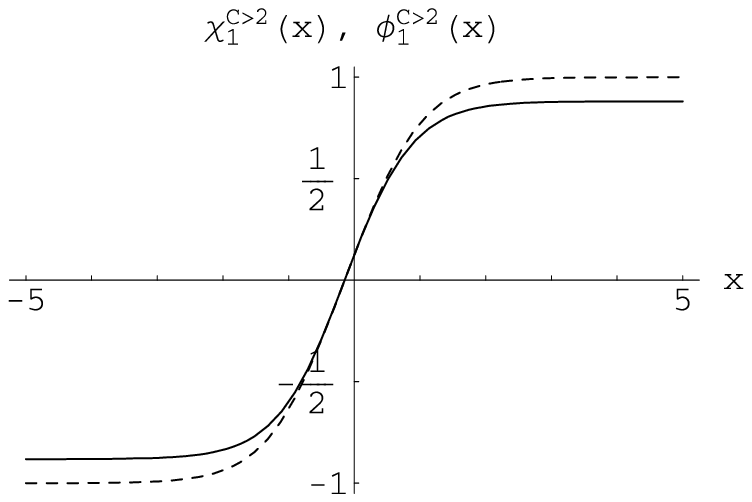}}
\quad
\subfigure{\includegraphics[height=40mm]{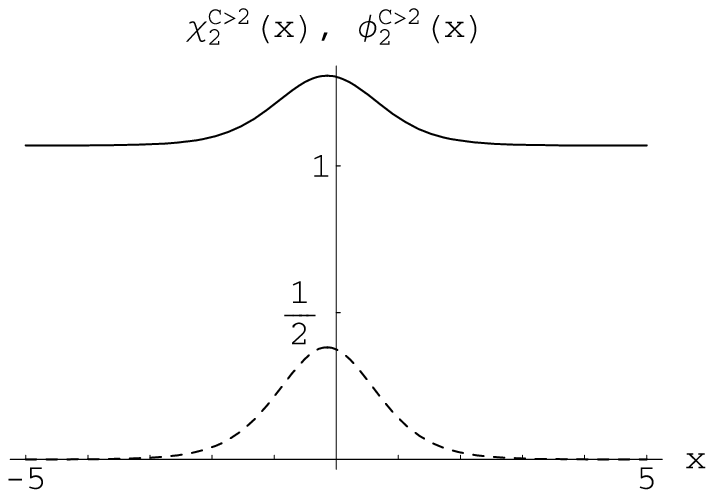}} }
\caption{Deformed (solid) and original (dotted) solutions for BNRT
model for $r=1$, $C_1=-1$ and $C=3$.} \label{r1solsC3}
\end{figure}

\medskip
For $C > 2$ we find that the profiles of $\chi_1$ and $\chi_2$ in (\ref{solsr1}) are kink and  lump-like respectively, as shown in Fig. \ref{r1solsC3}.

\bigskip

Before ending this section, let us recall that the values $r=1/4$ and $r=1,$ for which the integral (\ref{phi2C}) can be analyticaly implemented,
present special features: kinks of the $r={1}/{4}$ family are separatrix trajectories of a mechanical problem
which is Hamilton-Jacobi separable in parabolic coordinates, and for $r=1,$ the mechanical problem is separable in Cartesian coordinates;
these problems are known as Liouville type III and IV, respectivelly \cite{AMAJ}. However, even though the separable Liouville type problems
seem to lead to integrable expressions in equation (\ref{diecisiete}), in practice not all the kink orbits are algebraic, and this may preclude
the presence of analytical solutions.

\section{Comments and conclusions}

In this work we have presented a generalization of the deformation method, first introduced in \cite{Ba1}, which
allows to generate deformed potentials and solutions given a model of two real scalar field and its solutions.
The main new ingredient consist in the need of imposing a constraint on the functions used to deform the fields, required to preserve the relation between the original solutions, that live in orbits of the configuration space. As the construction of the deformed solutions involves the orbit constraint, the deformation must be implemented independently on each topological sector. Consequently, different orbits (even when belonging to the same sector)
can lead to different full deformed models and solutions.

The present version of the method applies to models with an associated superpotential $W(\chi_1,\chi_2).$ Although we have considered deformation functions depending on a single field, more general functions can be used, and reduced to the former case with the use of the orbit itself. Deformation functions depending on both scalar fields are now being considered in the context of models with holomorphic superpotentials \cite{AI2000}, as is the case of the bosonic sector of the (1+1)-dimensional N=2 SUSY Wess-Zumino model. We will further report on this possibility in future work.

As the described procedure is orbit dependent, whenever deforming a two-field model, the integrability of equation (\ref{diecisiete}) has to be analyzed separately in each case. However, in case of two-field theories which are associated to separable mechanical systems of the Liouville type, it seems that the present method will work very nicely, at least when one is restricted to consider algebraic kink orbits. Interesting examples of this kind are known, as the celebrated MSTB model \cite{MSTB} and other models proposed recently in Ref.~\cite{AJ}.

This new version of the deformation procedure provides a tool for studying more sophisticated systems.
It allows one to generate a diversity of new systems with their corresponding solutions,
which may contribute to improve the understanding of complex problems.

{\bf Acknowledgments:} VIA, DB and LL would like to thank CAPES, CNPq and PRONEX/CNPq/FAPESQ for partial support.
MAGL and JMG would like to thank the DGICYT and Junta de Castilla y Leon for partial financial support through the contracts: FIS2006-09417 and VAO13C05.



\begin{thebibliography}{99}
\addcontentsline{toc}{section}{References}
\bb{Raj}R. Rajaraman. {\it Solitons and instantons} (North-Holland, Amsterdam, 1982).
\bb{Vil}A. Vilenkin and E.P.S. Shellard, {\it Cosmic strings and other topological defects} (Cambridge, Cambridge/UK, 1994).
\bb{Man}N. Manton and P. Sutcliffe, {\it Topological solitons} (Cambridge, Cambridge/UK, 2004).
\bb{sugra}M. Cvetic and H.H. Soleng, Phys. Rep. 282, 159 (1997) [arXiv:hep-th/9604090].
\bb{brane}P.S. Wesson {\it Five-Dimensional Physics} (World Scientific, Singapore, 2006).
\bibitem{Ba1}D.~Bazeia, L.~Losano and J.M.C.~Malbouisson, 
              Phys.\ Rev.\ D {\bf 66}, 101701(R) (2002) [arXiv:hep-th/0209027].
\bibitem{BPS}M.K. Prasad and C.M. Sommerfield, Phys. Rev. Lett. {\bf 35}, 760 (1975); E.B. Bogomol´nyi, Sov. J. Nucl. Phys. {\bf24}, 449 (1976).
\bibitem{Ra1} R. Rajaraman, Phys. Rev. Lett. {\bf42}, 200 (1979).
\bibitem{Ba3} D.~Bazeia, W.~Freire, L.~Losano and R.F.~Ribeiro, 
              Mod.\ Phys.\ Lett.\ A {\bf 17}, 1945 (2002) [arXiv:hep-th/0205305].
\bibitem{trodden}M. Bowick, A. de Felice and M. Trodden, JHEP {\bf 0310}, 067 (2003) [arXiv:hep-th/0306224].
\bibitem{Ba4}C.A.~Almeida, D.~Bazeia, L.~Losano and J.M.C.~Malbouisson, 
             Phys.\ Rev.\ D {\bf 69}, 067702 (2004) [arXiv:hep-th/0405238];
             D. Bazeia and L. Losano, Phys. Rev. D {\bf73}, 025016 (2006) [arXiv:org/0511193];
             D. Bazeia, M.A. Gonzalez Leon, L. Losano and J. Mateos Guilarte, Phys. Rev. D {\bf73}, 105008 (2006) [arXiv:hep-th/0605127];
             V.I. Afonso, D. Bazeia, F.A. Brito, JHEP {\bf0608}, 073 (2006) [arXiv:hep-th/0603230].
\bb{dutra} A. de Souza Dutra and A.C. Amaro de Faría Jr, Phys. Rev. D {\bf72}, 087701 (2005);
           Phys. Lett. B {\bf642}, 274 (2006) [arXiv:hep-th/0610315].
\bb{gio}M. ~Giovannini, Phys. Rev. D {\bf74}, 087505 (2006)[arXiv:hep-th/0609136];
            Phys. Rev. D {\bf75}, 064023 (2007)[arXiv:hep-th/0612104].
\bb{guilarte} A. Alonso Izquierdo and J. Mateos Guilarte, Physica D {\bf220}, 31 (2006) [arXiv:nlin/0602059].
\bb{vacha} D.A. Steer and T. Vachaspati, Phys. Rev. D {\bf73}, 105021 (2006) [arXiv:hep-th/0602130].
\bb{morris}J.R. Morris, Phys. Rev. D {\bf51}, 697 (1995);
           Int. J. Mod. Phys. A {\bf13,} 1115 (1998) [arXiv: hep-ph/9707519].
\bb{ba}D. Bazeia and F. A. Brito, Phys. Rev. D {\bf61,} 105019 (2000) [arXiv:hep-th/9912015];
          Phys. Rev. D {\bf62,} 101701(R) (2000) [arXiv:hep-th/0005045];
          D. Bazeia, F.A. Brito and L. Losano, JHEP {\bf0611,} 064 (2006) [arXiv:hep-th/0610233].
\bb{sut}P. Sutcliffe, Phys. Rev. D {\bf68,} 085004 (2003) [arXiv:hep-th/0305198].
\bibitem{Gui2}A.A.~Izquierdo, M.A.~Gonzalez Leon and J.~Mateos Guilarte, Nonlinearity {\bf 13}, 1137 (2000) [arXiv:hep-th/0003224].
\bibitem{BNRT} D. Bazeia, M.J. dos Santos, R.F. Ribeiro,  Phys.\ Lett.\ A {\bf208}, 84 (1995) [arXiv:hep-th/0311265];
               D. Bazeia, J.R.S. Nascimento, R.F. Ribeiro, and D. Toledo, J. Phys. A {\bf30}, 8157 (1997) [arXiv:hep-th/9705224].
\bibitem{Gui1} A.A.~Izquierdo, M.A.~Gonzalez Leon and J.~Mateos Guilarte,
               Phys.\ Rev.\ D {\bf 65}, 085012 (2002) [arXiv:hep-th/0201200].
\bibitem{Abram}M.~Abramowitz and I.A.~Stegun, {\it Handbook of Mathematical Functions with Formulas,
Graphs, and Mathematical Tables} (Dover, New York, 1964).
\bb{AMAJ} A. Alonso Izquierdo, M. A. Gonzalez Leon, and J.
Mateos Guilarte, J. Phys. A {\bf31}, 209 (1998).
\bb{AI2000} A.~Alonso Izquierdo, M.~A.~Gonzalez Leon and J.~Mateos Guilarte,
            Phys.\ Lett.\  B {\bf 480}, 373 (2000) [arXiv:hep-th/0002082];
            D.~Bazeia, J.~Menezes and M.~M.~Santos, Phys.\ Lett.\  B {\bf 521}, 418 (2001) [arXiv:hep-th/0110111].
\bb{MSTB}C. Montonen, Nucl. Phys. B {\bf112}, 349 (1976);
          S. Sarker, S. Trullinger, and A. Bishop, Phys. Lett. A {\bf59}, 255 (1976).
\bb{AJ} A. Alonso Izquierdo and J. Mateos Guilarte, [arXiv:nlin/0611030].
\end{thebibliography}
\end{document}